\documentclass[prl,twocolumn,superscriptaddress]{revtex4}

\def\x{X} 	

\def\z{Z}

\def\H{H}
\def\W{W}

\def\w{\tilde{W}}

\usepackage{color}
\usepackage{amsfonts}
\usepackage{amsmath}
\usepackage{enumerate}
\usepackage{subfigure}
\usepackage{graphicx}

\begin{document}
\title{Topological Entanglement Entropy with a Twist}

\author{Benjamin J. Brown}
\email{benjamin.brown09@imperial.ac.uk}
\affiliation{Quantum Optics and Laser Science, Blackett Laboratory, Imperial College London, Prince Consort Road, London SW7 2AZ, UK}

\author{Stephen D. Bartlett} 
\affiliation{Centre for Engineered Quantum Systems, School of Physics, The University of Sydney, Sydney, NSW 2006, Australia}

\author{Andrew C. Doherty} 
\affiliation{Centre for Engineered Quantum Systems, School of Physics, The University of Sydney, Sydney, NSW 2006, Australia}

\author{Sean D. Barrett}
\thanks{Deceased 19 October 2012.}
\affiliation{Quantum Optics and Laser Science, Blackett Laboratory, Imperial College London, Prince Consort Road, London SW7 2AZ, UK}

\begin{abstract}
Defects in topologically ordered models have interesting properties that are reminiscent of the anyonic excitations of the models themselves. For example, dislocations in the toric code model are known as \emph{twists} and possess properties that are analogous to Ising anyons.  We strengthen this analogy by using the topological entanglement entropy as a diagnostic tool to identify properties of both defects and excitations in the toric code.  Specifically, we show, through explicit calculation, that the toric code model including twists and dyon excitations has the same quantum dimensions, the same total quantum dimension, and the same fusion rules as an Ising anyon model.  \end{abstract}

\maketitle

A fascinating class of many-body quantum systems are those that exhibit topological order~\cite{Wentop}.
Such systems are characterized by a gapped ground state manifold, with a degeneracy that depends on the boundary conditions, and have anyonic quasi-particle excitations. The degeneracy is robust to local perturbations, and therefore such systems are promising candidates for storing and manipulating quantum information \cite{Dennis, Kitaev, Freedman, Nayak}.

The structure of topologically-ordered systems can be further enriched by the use of domain walls or defects, across which quasi-particles transform nontrivially.  We will be most interested in studying these defects in explicit lattice models~\cite{Bombin,Beigi,Kong,YouWen1,YouWen2}, and in particular the model introduced by Bombin~\cite{Bombin}. Essentially the same defects have also been studied in Chern-Simons theories~\cite{BarkeshliWen,BarkeshliQi,Barkeshli}. In both of these settings, it has been shown that such defects can be viewed as having anyonic-like properties that are not associated with the underlying model~\cite{BarkeshliWen,Bombin,BarkeshliQi,YouWen1,Barkeshli}.  
More specifically, in a particular two-dimensional topologically ordered spin lattice model known as the toric code, such a lattice dislocation leads to interesting behaviour: the points where dislocations terminate, known as twists, interact with the anyons of the toric code to reproduce properties, such as fusion rules, of nonabelian (specifically, Ising) anyons~\cite{Bombin}.  
This is particularly surprising, as all excitations of the toric code are abelian anyons.  

In this paper, we further interrogate the analogy between twists and Ising anyons by using topological entanglement entropy (TEE)~\cite{Preskill, Levin} as a diagnostic.  
Specifically, for a toric code model containing twists, we use the TEE to determine: (i) the total quantum dimension of the lattice with twists; (ii) the quantum dimensions of the objects (quasi-particles and defects) on the lattice, and (iii) the quantum dimensions of all of their fusion products, allowing us to reconstruct the fusion rules for twists and excitations.  Our results coincide precisely with those of the Ising anyon model, lending further support to the analogy between the latter and the toric code with twists.  We note that TEE is a particularly useful quantity in the context of the toric code, as both the parent Hamiltonian and the modified one with twists are described within the stabilizer formalism and so allow the TEE to be calculated exactly. Our results show that defects in topological lattice models, which are of considerable interest because of their exotic anyonic properties, can be probed using the topological entanglement entropy.


We first review some relevant properties of topological order and anyons.  
A topologically-ordered system with fixed locations of a fixed number of anyons is described by a Hilbert space $\mathcal{H}_{\text{topo}}$. 
The dimension of $\mathcal{H}_{\text{topo}}$ for a system with $N$ type-$a$ anyons in the limit of large $N$ is given by $\text{dim}\ \mathcal{H}_{\text{topo}} \approx  (d_a)^N/ \mathcal{D}^2 $. Here, $d_a$ is the \emph{quantum dimension} of a type-$a$ anyon, not necessarily an integer, and $\mathcal{D} = \sqrt{\sum_x d_x^2}$ is the \emph{total quantum dimension}, where the summation is over all the anyon types.
Two anyon models are relevant to our study:  $\mathbb{Z}_2$ abelian anyons, and Ising anyons (see, e.g., Ref.~\cite{KitaevHoney}).  The elementary excitations of a $\mathbb{Z}_2$ abelian anyon model are the electric charge, $e$, the magnetic vortex, $m$, the dyon, $\epsilon$ and the vacuum, $1$; all have quantum dimension $d_a = 1$ and so $\mathcal{D}_{\mathbb{Z}_2} =  2$.  A model possessing Ising anyons includes: an Ising anyon $\sigma$, a fermion $\psi$, and the vacuum $1$, where $d_1 = d_\psi = 1$ and $d_\sigma = \sqrt{2}$.  These anyons are nonabelian, as when two $\sigma$ anyons fuse, there are two possible fusion outcomes $1$ and $\psi$.  Like the $\mathbb{Z}_2$ quantum double, this model has $\mathcal{D}_{\text{Ising}} = 2$.


Entropy is a powerful tool for probing topological properties of a model. For the ground states of  quantum many-body systems with local Hamiltonians the von Neumann entropy $S_A$ of the reduced state on a region $A$ is expected to satisfy an area law \cite{Wolf, Eisert}. Restricting to two-dimensional systems, the von Neumann entropy of a region $A$ of a topologically ordered system is expected to have a universal correction $\gamma$ to $S_A$ such that \cite{Preskill, Levin}
\begin{equation}
S_A = \alpha L_A - n_A \gamma, \label{vN}
\end{equation}
where $L_A$ is the length of the boundary of region $A$ (which should be smooth), $n_A$ is the number of disconnected boundaries enclosing region $A$ and $\alpha$ is a constant that depends on the microscopic properties of the Hamiltonian. 

\begin{figure}
\subfigure{
\includegraphics[scale=5]{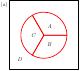}
\label{mercedes}
}
\subfigure{
\includegraphics[scale=5]{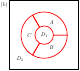}
\label{annulus}
}
\caption{\subref{mercedes} Regions used to calculate topological entanglement entropy. 
\subref{annulus} Regions used to probe the quantum dimension of an anyon on region $D_1$. 
}
\end{figure}

For a given topological model, the total quantum dimension is related to $\gamma$ via $\gamma = \log \mathcal{D}$ \cite{Preskill, Levin}.  As such, the von Neumann entropy can be used to probe the anyonic properties of the system.  It is useful to eliminate the the boundary contribution of Eq.~(\ref{vN}) by computing a linear combination of the von Neumann entropies of overlapping regions as in Fig.~\ref{mercedes} to obtain the \emph{topological entanglement entropy}
\begin{equation}
  S_{\text{topo}} = S_A + S_B +S_C - S_{AB} -S_{BC} - S_{CA} + S_{ABC} \,, \label{topo}
\end{equation}
where $S_R$ is the von Neumann entropy of region $R$~\cite{Preskill, Levin}.  The TEE is a function solely of the total quantum dimension of the model, $S_{\text{topo}}=-\gamma = -\log \mathcal{D}$.

Additional information about the anyonic properties of the model can be revealed by defining analogous entropic quantities using regions with noncontractable boundaries~\cite{Zhang}, as in Fig.~\ref{annulus}.  By using excited states rather than the ground state, our entropic calculations will depend explicitly on the number of anyons present and their placement within the regions.  Consider a state describing an anyon of type $a$ on a simply-connected region, labeled $A$;  the presence of the anyon gives a correction to the von Neumann entropy as $S_A(a) = \alpha L_A - \log(\mathcal{D}/d_a)$ \cite{Preskill, Dong}.  Again, the boundary contribution can be removed with a suitable combination of the entropies on a nontrivial partition; a suitable choice is an annular region shown in Fig.~\ref{annulus} defining a disconnected region $D = D_1 \cup D_2$.  For a state describing a pair of anyons of type $a$ that fuse to the vacuum, with one in region $D_1$ and the other in region $D_2$, the entropy
$S_{\text{ann}}(a) = S_A + S_B +S_C - S_{AB} -S_{BC} - S_{CA} + S_{ABC}$ gives a value 
\begin{equation}
  S_{\text{ann}}(a) = -2\log{\mathcal{D}/d_a}\,, \label{ann}
\end{equation}
where the factor of $2$ arises from the boundary connectivity as in Eq.~(\ref{vN}).

In this work we propose to use $S_{\text{ann}}$ to infer fusion rules for both anyons and twists. For this purpose we must determine how  $S_{\text{ann}}$ behaves in the situation where there are multiple anyons in region $D_1$.   If $\mathcal{H}_{\text{topo}}$ is in a state where the anyons $a_1, a_2,\ldots$ in $D_1$ fuse to a definite fusion product $x_1$, then $S_{\text{ann}}(a_1,a_2,\ldots) = S_{\text{ann}}(x_1)$.   
However, in general, anyons can fuse through multiple channels, with the probability (amplitude) of a given fusion outcome determined by the state encoded in $\mathcal{H}_{\text{topo}}$.
Considering only states where the connected region $ABC$ as well as the entire lattice fuse to the vacuum, the outcome $x_1$ determines explicitly $x_2$, the fusion product of the anyons on $D_2$. In this important special case, we can use results from Ref.~\cite{Dong} to compute $S_{\text{ann}}$ and we find the explicit formula
\begin{equation}
S_{\text{ann}}(\{x\}) = - 2 \log{\mathcal{D}} - \sum_x P_x  \log [P_x/(d_{x_1} d_{x_2})]\,, \label{anntop}
\end{equation}  
where the sum is over all fusion outcomes $x = (x_1,x_2)$, each of which occur with probability $P_x$.  This equation provides the principal diagonostic we will use in this paper to characterise the anyonic properties of a model's quasi-particles and defects.  As we will show, the TEE of a model with twists using annular regions with a noncontractable boundary will yield identical results to the analogous entropic calculation using excited states of a model with Ising anyons in place of twists.


\begin{figure}
\includegraphics[scale=1.5]{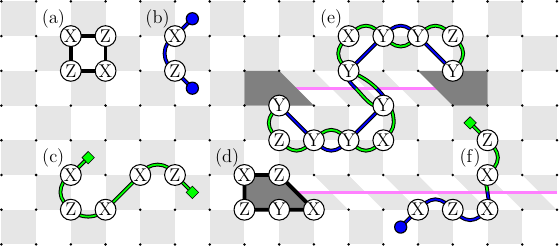}
\caption{ Pauli $X$, $Y$ and $Z$ operators are shown in black circles.  (a) A four-body interaction term for Kitaev's toric code Hamiltonian. (b) and (c) Strings of Pauli operators which respectively generate electric and magnetic charged pairs at the end points. Electric(magnetic) strings can only be generated on grey(white) plaquettes.  (d) The five-body interaction term for the twist plaquette. (e) A logical operator in the form of a dyon chain terminating at two twists. (f) A Pauli string operator that cross a domain wall.}\label{Twist}
\end{figure}

The specific model we consider here is the $\mathbb{Z}_2$ quantum double, also known as Kitaev's toric code model~\cite{Kitaev}, and we focus on a formulation due to Wen~\cite{Wenmodel, Wenmanybody}.  On a square lattice with spins on the vertices, labeled $(j,k)$, the model is described by the Hamiltonian $\H = - \sum_{j,k} \W_{j,k}$ with $\W_{j,k} = \x_{j,k} \z_{j+1,k} \z_{j,k+1} \x_{j+1,k+1}$, shown on the lattice in Fig.~\ref{Twist}(a).  We note that $\H$ is a stabilizer Hamiltonian, and all term commute. 
The low energy excitations of this model are $\mathbb{Z}_2$ anyons, which are created and annihilated at the end points of string operators, shown at Fig~\ref{Twist}(b) and (c).
Non-trivial domain walls can be introduced in this model by shifting the geometry of the lattice along a line, such that we have even(odd) plaquettes adjacent to other even(odd) plaquettes~\cite{Bombin, Kong}. We show such dislocations along pink lines in Fig.~\ref{Twist}. The interaction terms of the Hamiltonian along the dislocation then change to correspond with the new geometry, but remain locally equivalent to the old interaction terms except at the ends of the dislocation. The five body interaction terms where dislocations terminate, shown at Fig.~\ref{Twist}(d), are known as \emph{twists}~\cite{Bombin}. We note that the insertion of twists into the toric code model changes the ground state degeneracy, as can be seen from the fact that the dyon chain stabilizers shown in Fig.~\ref{Twist}(e) commute with the Hamiltonian.

The presence of twists in the toric code model prevents us from globally defining the electric or magnetic charge of an anyon:  an $e(m)$ anyon that crosses a domain wall will transform into an $m(e)$ anyon, as shown in Fig.~\ref{Twist}(f), although the dyons remain well defined.  In addition, the twists at the end points of these dislocations can be used to reproduce some of the fusion rules of the Ising anyon model \cite{Bombin}.  This analogy is achieved by considering only the vacuum, toric code dyons $\epsilon$, and one type (handedness) of twist.  By treating $\epsilon$ dyons as the fermions ($\psi$-type anyons) of an Ising anyon model, and the twists as $\sigma$-type anyons, one can reproduce the fusion properties of the Ising anyon model. In this sense, twists exhibit anyon-like behaviour.  
Here we use TEE as a diagnostic tool to further interrogate this analogy.


We can exploit the stabilizer formalism to calculate von Neumann entropies for regions of the vacuum and excited states of these topological models, using the method of Ref.~\cite{Fattal}.  
This method makes use of a canonical form of a generating set of the stabilizer group defined by a bipartition of the lattice into regions $A$ and $\overline{A}$. The stabilizer generators can be grouped into those entirely in $A$, those entirely in $\bar{A}$ and those that have support on both partitions. For the set of stabilizer generators that cross the bipartition, a canonical form can be constructed such that the local support of the generators $K_i$ on $A$, denoted by $\tilde{K}_i$, either commutes with all other $\tilde{K}_j$, or anti-commutes with only one other such element $\tilde{K}_{i'}$. These pairs of stabilizer generators that anti-commute when restricted to $A$ describe a pair of maximally entangled qubits sitting across the boundary.  Counting these pairs gives the von Neumann entropy $S_A$.  Applying this method to the toric code vacuum~\cite{Hamma} gives $S_A = \frac{1}{2}L_A - 1$, and therefore $S_{\text{topo}}=-1$ and $\mathcal{D}=2$ as expected for a model with $\mathbb{Z}_2$ anyons. (Here we choose the length of the boundary $L_A$ to correspond to the number of plaquettes that cross the boundary.) We note that the following results are independent of the topology of the lattice, because homologically non-trivial stabilizers of a topologically non-trivial model can always be deformed away from the regions considered in the present entropy calculations~\cite{Hamma}.  With this in mind, we obtain the same results for the following calculations regardless of the boundary conditions of the lattice. 


\begin{figure}
\includegraphics[scale=1.5]{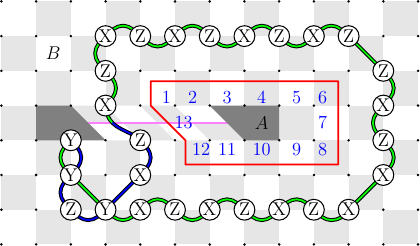}
\caption{\label{ent} The boundary $\partial A$ shown in red of a simply connected region enclosing a twist on a lattice. We number the cut plaquettes with a single index shown in blue, here, $L_A=13$. We also show a logical operator deformed outside region $A$. }
\end{figure}


We first consider a vacuum state of a toric code model with some number of twists, and calculate the von Neumann entropy for a simply connected region enclosing one twist as shown in Fig.~\ref{ent}.  To do this calculation, we must identify a generating set of stabilizers for a vacuum state, in the canonical form described above.  Details of this calculation are presented explicitly in the Supplementary Material~\cite{supp}, and all subsequent calculations described below are straightforward variations of this calculation.  We find that $S_A = \frac{1}{2}L_A - \frac{1}{2}$, as expected for a region of an Ising anyon model containing an $\sigma$-type Ising anyon with $d_\sigma = \sqrt{2}$.    

This method allows us to calculate the effect of twists on the TEE of the vacuum, using Eq.~(\ref{topo}) and regions defined as in Fig.~\ref{mercedes}.  We find that for a vacuum state of a model with any number of twists, the TEE is $S_{\text{topo}} = - 1 $ independent of the locations of the twists.  This outcome corresponds to a model with total quantum dimension $\mathcal{D}=2$, consistent with models with either Ising or $\mathbb{Z}_2$ abelian anyons.

We next show that, using the annular regions with non-connected boundaries as in Fig.~\ref{annulus} which can probe the properties of anyons, the entropy $S_{\text{ann}}(1)$ for annular region of the vacuum state of a model containing twists is equivalent to the entropy $S_{\text{ann}}(\sigma)$ of a state of an Ising anyon model with $\sigma$-type Ising anyons.  We use a partitioning as in Fig.~\ref{annulus}, with a twist on region $D_1$ and another on $D_2$, as shown in Fig.~\ref{twotwist}. We achieve canonical form for region $ABC$ using the method described in the Supplementary Material~\cite{supp}, but now with disconnected boundaries we must consider the canonical form of the stabilizer generators for each boundary separately.  As above, any dyon string stabilizer that crosses region $ABC$ can be deformed entirely onto region $D_2$.  Calculating the entropy for the various regions, we find $S_{ABC} = \frac{1}{2} L_{ABC} - 1$, with all other regions giving $S_R = \frac{1}{2}L_R - 1$, and so $S_{\text{topo}} = - 1 $.   We compare this result to a calculation of $S_{\text{ann}}(\sigma)$ for a state of an Ising anyon model with a $\sigma$-type Ising anyon in each region $D_1$ and $D_2$, using Eq.~(\ref{ann}), and obtain an identical result, $S_{\text{ann}}(\sigma)=-1$.  Thus, from an entropic perspective, the twists appear as $\sigma$-type Ising anyons.  A similar calculation involving a $\epsilon$ quasi-particle on region $D_1$ and another on region $D_2$ gives $S_{\text{ann}}(\epsilon)=S_{\text{ann}}(\psi)=1$ as expected from a Ising fermion $\psi$ in the Ising anyon model.

\begin{figure}

\subfigure{
\includegraphics[scale=3.5]{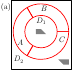}
\label{twotwist}
}
\subfigure{
\includegraphics[scale=3.5]{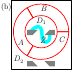}
\label{fourztwist}
}
\subfigure{
\includegraphics[scale=3.5]{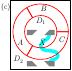}
\label{fourxtwist}
}

\caption{\subref{twotwist} The annular topological entanglement entropy regions with one twist on region $D_1$ and one twist on region $D_2$.  \subref{fourztwist} The lattice configuration where two twists lie on $D_1$ and two lie on $D_2$, in an eigenstate of $\bar{Z}$ (defined as a dyon chain stabilizer as in Fig.~\ref{Twist}(e)).  \subref{fourxtwist} A configuration with four twists in an eigenstate of $\bar{X}$. }
\end{figure}


Finally, we use TEE to probe the fusion rules associated with quasi-particles and twists.  We show that pairs of twists fuse according to the more sophisticated nonabelian fusion rules of two $\sigma$-type Ising anyons, and modify the TEE according to our Eq.~(\ref{anntop}) in a way that depends on the specific fusion channel. We place two twists on region $D_1$ and two twists on region $D_2$.  The channel by which the two twists on region $D_1$ fuse together is dependent on the logical state, as characterised by the logical operators that are given by dyon strings between twists.  

We first consider the lattice prepared in an eigenstate of the logical-$Z$ operator $\bar{Z}$ shown in Fig.~\ref{fourztwist}. The $+1$ eigenstate of $\bar{Z}$ corresponds to two $\sigma$-type anyons on region $D_1$ fusing to the vacuum $1$ with certainty, whereas the $-1$ eigenstate fuses to the fermion $\psi$ with certainty.  In either case, we find $S_{\text{ann}}(\{(\sigma,\sigma)\rightarrow 1\}) = S_{\text{ann}}(\{(\sigma,\sigma)\rightarrow \psi\}) = - 2$. Comparing this with Eq.~(\ref{anntop}), we have the result we anticipate from the Ising anyon model. 


We next consider the case where the lattice is prepared in an eigenstate of the logical-$X$ operator $\bar{X}$ shown in Fig.~\ref{fourxtwist}. This is analogous to the Ising anyon model where one $\sigma$-type anyon on region $D_1$ and one $\sigma$-type anyon on region $D_2$ will fuse to give one fusion outcome with certainty. However, fusing the two particles on region $D_1$ can produce the vacuum $1$ or a fermion $\psi$ with equal probability $P_{(1,1)} = P_{(\psi,\psi)} = \frac{1}{2}$.  The details of the TEE calculation are more involved in this case, as the operator $\bar{X}$ cannot be deformed away from region $ABC$. A simple modification of the above calculations gives the result $S_{ABC} = \frac{1}{2} L_{ABC} - 1$; full details of this calculation can be found in the Supplementary Material~\cite{supp}.
The other regions used fuse to vacuum, giving $S_{\text{ann}}(\{(\sigma,\sigma)\rightarrow 1\ \text{or}\ \psi\}) = - 1$, which from Eq.~(\ref{anntop}) is the result we expect from the analogous Ising anyon case. 
 

To complete the analogy, one might seek to identify the braiding properties of twists, by using the techniques of Ref.~\cite{Zhang} for example, and compare them to Ising anyons. However, as noted by Bombin~\cite{Bombin}, braiding of twists are not expected to be well defined and we have not considered them in this work.  We note that it may be possible to define robust braiding operations up to an overall phase~\cite{YouWen2,Barkeshli,Vaezi}, and in particular one might seek to use techniques from Ref.~\cite{Barkeshli} to check explicitly that the twists in this model have the expected braiding properties.  We leave this for future work.

We note that, although the TEE is defined using the von Neumann entropy, the results of Ref.~\cite{Flammia} show that all topological Renyi entropies are equivalent.  The equivalence between all Renyi entropies for the topological properties calculated here is easily seen through our stabilizer calculation, as the entanglement spectrum arising from the reduced density matrices corresponds to an integer number of maximally entangled Bell pairs, and is always uniform.  As a result, the diagnostic topological properties studied here can be directly measured (i.e., they correspond to observables) using the 2-Renyi entropy and the methods proposed in Ref.~\cite{Demler}.

In conclusion, our results demonstrate how the TEE can be used as a diagnostic to probe the anyonic-like properties of defects in topological models.  This new tool may assist in the construction of topological lattice models with defects demonstrating richer anyonic properties than the underlying model, with potential applications to topological quantum computing.

We acknowledge H. Bombin, C. Brell and J. Wootton for useful discussions. This work is supported by the EPSRC and the ARC via the Centre of Excellence in Engineered Quantum Systems (EQuS), project number CE110001013.

\newpage

\appendix

\section{Supplementary Material}

In this Supplementary Material, we provide details of the stabilizer calculation for the von Neumann entropy for various regions in the presence of twists. We will pay particular attention to the simplest case of the entropy of a simply connected region enclosing a twist, the other cases involve simple modifications of this argument. As noted in the main text the calculation applies the general methods of~\cite{Fattal}. We assume that the reader is familiar with the stabiliser formalism for describing quantum states, for more details on this we refer the reader to~\cite{Fattal} and references therein.

\subsection*{Von Neumann entropy of a region enclosing a twist}
Consider a simply connected region of the vacuum state of a toric code enclosing one twist.  We will show that $S_A = \frac{1}{2}L_A - \frac{1}{2}$, as expected for a region of an Ising anyon model containing an $\sigma$-type Ising anyon with $d_\sigma = \sqrt{2}$.  To do this calculation, we must identify a generating set of stabilizers for a vacuum state in the canonical form.  

We consider a specific example of a lattice containing a single twist as follows:
\begin{center}
\includegraphics[scale=2]{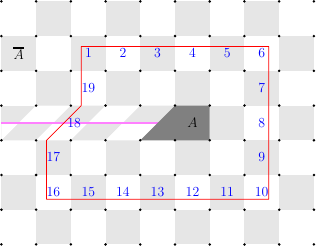}
\end{center}
with a partition $\partial A$ (shown in red).  Although there can exist stabilizers that take the form of dyon strings terminating at the twist, we can multiply by local stabilizers to deform such strings such that they have no support on $A$.  These deformed strings are indicated in Figure 3 of the main text. Thus, a complete set of stabilizer generators that cross the partition $\partial A$ is given by the plaquette operators $\W_l$ for $1 \le l \le L_A $, above.  For example, the plaquette operator corresponding to $l=4$ is:
\begin{center}
\includegraphics[scale=0.75]{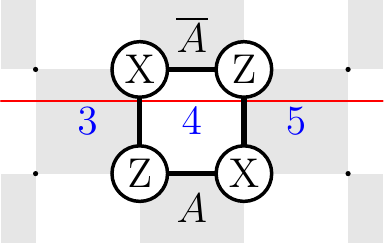}
\end{center}
We define $\w_l$ to be the support of these operators on $A$.  For example, $\w_4$ has the form:
\begin{center}
\includegraphics[scale=0.75]{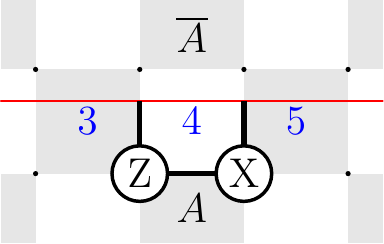}
\end{center}
The $\W_l$ operators are not in canonical form, as each $\w_l$ anti-commutes with both $\w_{l-1}$ and $\w_{l+1}$.  In our example, $\w_4$ anti-commutes with the following two operators $\w_{3}$ and $\w_{5}$ given by: 
\begin{center}
\includegraphics[scale=0.75]{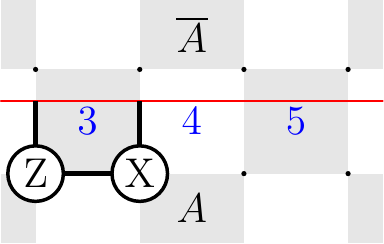} \qquad \includegraphics[scale=0.75]{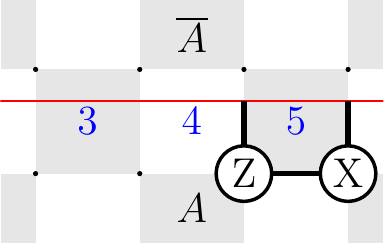}
\end{center}
To achieve canonical form, we will replace one of the stabilizer generators on the boundary with a stabilizer generator confined to $\bar{A}$, as follows.  First, consider the product of all the stabilizer generators $\W_{\partial A} = \prod_{l\in\partial A} \W_l$ that lie on the boundary.  In our example, $\W_{\partial A}$ takes the form:
\begin{center}
\includegraphics[scale=2]{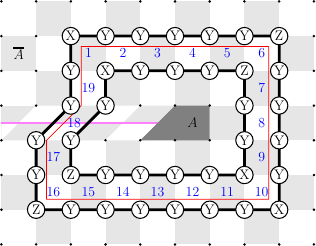}
\end{center}
Notice that $\W_{\partial A}$ commutes with all operators $\w_l$.  Also, consider the product $\W_A = \prod_{i\in A} \W_i$ of all stabilizer generators contained entirely in $A$.  In our example, this product $\W_A$ takes the form
\begin{center}
\includegraphics[scale=2]{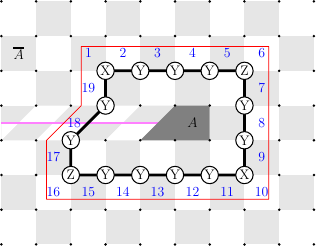}
\end{center}
The product of $\W_{\partial A}$ and $\W_A$ gives an operator supported entirely on $\bar{A}$, as follows:
\begin{center}
\includegraphics[scale=2]{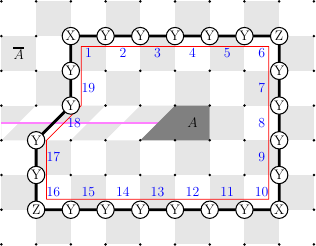}
\end{center}
(Note that, in the untwisted toric code, one can construct two such operators using the natural partition of toric code stabilizers into two commuting subsets.  With the presence of a twist, however, there is only a single such operator.)  

We will now modify our set of stabilizer generators, by removing one of the generators on the boundary (to be specific, say $\W_1$) and replacing it with the product of $\W_{\partial A}$ and $\W_A$.  It should be clear from the preceding argument that this replacement is possible because $\W_1$ can be expressed as a product of our new stabilizer generator and the generators in $A$ and $\partial A$.  

We now will modify the stabilizer generators on the boundary into a new set $K_l$ that has canonical form, as follows.  We retain the stabilizer generators $K_{l=2j} = \W_{2j}$ for integer $j$.  For these stabilizers, the operators $\tilde{K}_{l=2j}$ obtained by restricting $K_{l=2j}$ to region $A$ take a form as shown below for the particular example of $l=12$:
\begin{center}
\includegraphics[scale=2]{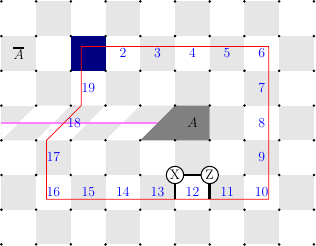}
\end{center}
We replace $\W_{2j+1}$ with the product $K_{2j+1} = \prod_{i=2j+1}^{L_A} \W_i$.  (Again, it is clear that $\W_{2j+1}$ can be expressed as a product of the new stabilizer generators $K_l$.)  For these new stabilizer generators, the operators $\tilde{K}_{2j+1}$ take a form as shown below for the particular example of $l=13$:
\begin{center}
\includegraphics[scale=2]{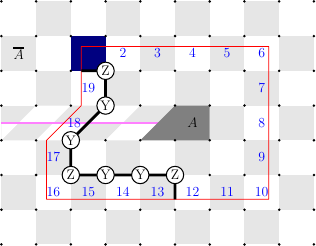}
\end{center}
Clearly, we have that, for all $j$, the new operators $\tilde{K}_{2j}$ and $\tilde{K}_{2j+1}$ anti-commute.  It can also be seen that the operator $\tilde{K}_{2j-1}$, shown as follows for $\tilde{K}_{11}$:
\begin{center}
\includegraphics[scale=2]{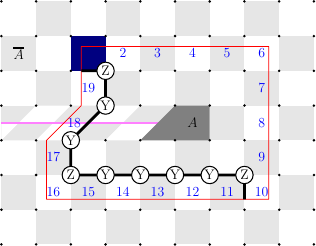}
\end{center}
clearly commutes with $\tilde{K}_{2j}$ (that is, $\tilde{K}_{11}$ commutes with $\tilde{K}_{12}$, by inspection), and thus we have achieved the required canonical form.   
Counting the pairs of locally anti-commuting generators $\tilde{K}_{2j}$ and $\tilde{K}_{2j+1}$ for $j=1,\ldots, L_A-1$, there are now $(L_A-1)/2$ such pairs, and so we have the stated result $S_A = \frac{1}{2} L_A - \frac{1}{2}$. 

\subsection*{Von Neumann entropy of an annular region with various arrangements of twists.}
We also needed to calculate von Neumann entropies for annular regions ABC with twists outside the annulus as indicated in Figure 4 of the main paper. This means that the boundary of the region of interest ABC is now composed of two disjoint pieces. However the modifications of the argument above are nontrivial only in the case indicated in figure 4c of the main paper where the lattice is prepared in an eigenstate of the logical-$X$ operator $\bar{X}$. This is analogous to the Ising anyon model where one $\sigma$-type anyon on region $D_1$ and one $\sigma$-type anyon on region $D_2$ will fuse to give one fusion outcome with certainty. However, fusing the two particles on region $D_1$ can produce the vacuum $1$ or a fermion $\psi$ with equal probability $P_{(1,1)} = P_{(\psi,\psi)} = \frac{1}{2}$.  The details of the TEE calculation are more involved in this case, as the operator $\bar{X}$ cannot be deformed away from region $ABC$.  

We therefore require a different method to find a canonical form for region $ABC$. Denoting the two connected segments of the boundary by $\partial ABC_\beta$, for $\beta =1,2$, we rewrite the indices of the plaquettes that cut these boundaries as $ \W_{\beta, l} $, with $1 \le l \le L_\beta$ numbering the generators around the boundary as before.
We deform $\bar{X}$ such that it only has common support with the generators $\W_{1,L_1-1}$, $\W_{1,L_1}$, $\W_{2,L_2-1}$, and $\W_{2,L_2}$ on the boundaries.  To achieve canonical form, we replace the generators 
\begin{align}
  \W_{1,L_1-1} &\rightarrow \prod_{\partial ABC_1} \W_{1,l} \,, \\
  \W_{2,L_2-1} &\rightarrow \prod_{\partial ABC_2} \W_{2,l} \,, \\ 
  \W_{1,L_1} &\rightarrow \prod_{\substack{\partial ABC_1\\ k=\text{even}}} \W_{1,k}\prod_{\substack{\partial ABC_2\\ l=\text{even}}}\W_{2,l}
\end{align}  
such that their local support commutes with the local support all the other generators. The remaining cut generators, including $\bar{X}$, are then transformed such that their local supports pairwise anti-commute, giving the required canonical form. This gives the result $S_{ABC} = \frac{1}{2} L_{ABC} - 1$.


\begin{thebibliography}{20}
\expandafter\ifx\csname natexlab\endcsname\relax\def\natexlab#1{#1}\fi
\expandafter\ifx\csname bibnamefont\endcsname\relax
  \def\bibnamefont#1{#1}\fi
\expandafter\ifx\csname bibfnamefont\endcsname\relax
  \def\bibfnamefont#1{#1}\fi
\expandafter\ifx\csname citenamefont\endcsname\relax
  \def\citenamefont#1{#1}\fi
\expandafter\ifx\csname url\endcsname\relax
  \def\url#1{\texttt{#1}}\fi
\expandafter\ifx\csname urlprefix\endcsname\relax\def\urlprefix{URL }\fi
\providecommand{\bibinfo}[2]{#2}
\providecommand{\eprint}[2][]{\url{#2}}

\bibitem[{\citenamefont{Wen}(1989)}]{Wentop}
\bibinfo{author}{\bibfnamefont{X.-G.} \bibnamefont{Wen}},
  \bibinfo{journal}{Phys. Rev. B} \textbf{\bibinfo{volume}{40}},
  \bibinfo{pages}{7387} (\bibinfo{year}{1989}).

\bibitem[{\citenamefont{Dennis et~al.}(2002)\citenamefont{Dennis, Kitaev,
  Landahl, and Preskill}}]{Dennis}
\bibinfo{author}{\bibfnamefont{E.}~\bibnamefont{Dennis}},
  \bibinfo{author}{\bibfnamefont{A.}~\bibnamefont{Kitaev}},
  \bibinfo{author}{\bibfnamefont{A.}~\bibnamefont{Landahl}}, \bibnamefont{and}
  \bibinfo{author}{\bibfnamefont{J.}~\bibnamefont{Preskill}},
  \bibinfo{journal}{J. Math. Phys.} \textbf{\bibinfo{volume}{43}},
  \bibinfo{pages}{4452} (\bibinfo{year}{2002}).

\bibitem[{\citenamefont{Kitaev}(2003)}]{Kitaev}
\bibinfo{author}{\bibfnamefont{A.}~\bibnamefont{Kitaev}},
  \bibinfo{journal}{Ann. Phys.} \textbf{\bibinfo{volume}{303}},
  \bibinfo{pages}{2} (\bibinfo{year}{2003}).

\bibitem[{\citenamefont{Freedman et~al.}(2002)\citenamefont{Freedman, Kitaev,
  Larsen, and Wang}}]{Freedman}
\bibinfo{author}{\bibfnamefont{M.~H.} \bibnamefont{Freedman}},
  \bibinfo{author}{\bibfnamefont{A.}~\bibnamefont{Kitaev}},
  \bibinfo{author}{\bibfnamefont{M.~J.} \bibnamefont{Larsen}},
  \bibnamefont{and} \bibinfo{author}{\bibfnamefont{Z.}~\bibnamefont{Wang}},
  \bibinfo{journal}{Bull. Am. Math. Soc.} \textbf{\bibinfo{volume}{40}},
  \bibinfo{pages}{31} (\bibinfo{year}{2002}).

\bibitem[{\citenamefont{Nayak et~al.}(2008)\citenamefont{Nayak, Simon, Stern,
  Freedman, and Sarma}}]{Nayak}
\bibinfo{author}{\bibfnamefont{C.}~\bibnamefont{Nayak}},
  \bibinfo{author}{\bibfnamefont{S.~H.} \bibnamefont{Simon}},
  \bibinfo{author}{\bibfnamefont{A.}~\bibnamefont{Stern}},
  \bibinfo{author}{\bibfnamefont{M.}~\bibnamefont{Freedman}}, \bibnamefont{and}
  \bibinfo{author}{\bibfnamefont{S.~D.} \bibnamefont{Sarma}},
  \bibinfo{journal}{Rev. Mod. Phys.} \textbf{\bibinfo{volume}{80}},
  \bibinfo{pages}{1083} (\bibinfo{year}{2008}).

\bibitem[{\citenamefont{Bombin}(2010)}]{Bombin}
\bibinfo{author}{\bibfnamefont{H.}~\bibnamefont{Bombin}},
  \bibinfo{journal}{Phys. Rev. Lett.} \textbf{\bibinfo{volume}{105}},
  \bibinfo{pages}{030403} (\bibinfo{year}{2010}).

\bibitem[{\citenamefont{Beigi et~al.}(2011)\citenamefont{Beigi, Shor, and
  Whalen}}]{Beigi}
\bibinfo{author}{\bibfnamefont{S.}~\bibnamefont{Beigi}},
  \bibinfo{author}{\bibfnamefont{P.}~\bibnamefont{Shor}}, \bibnamefont{and}
  \bibinfo{author}{\bibfnamefont{D.}~\bibnamefont{Whalen}},
  \bibinfo{journal}{Commun. Math. Phys.} \textbf{\bibinfo{volume}{306}},
  \bibinfo{pages}{663} (\bibinfo{year}{2011}).

\bibitem[{\citenamefont{Kitaev and Kong}(2012)}]{Kong}
\bibinfo{author}{\bibfnamefont{A.}~\bibnamefont{Kitaev}} \bibnamefont{and}
  \bibinfo{author}{\bibfnamefont{L.}~\bibnamefont{Kong}},
  \bibinfo{journal}{Commun. Math. Phys.} \textbf{\bibinfo{volume}{313}},
  \bibinfo{pages}{351} (\bibinfo{year}{2012}).
  
\bibitem[{\citenamefont{You and Wen}(2012)}]{YouWen1}
\bibinfo{author}{\bibfnamefont{Y.-Z.} \bibnamefont{You}} \bibnamefont{and}
  \bibinfo{author}{\bibfnamefont{X.-G.}~\bibnamefont{Wen}},
  \bibinfo{journal}{Phys. Rev. B} \textbf{\bibinfo{volume}{86}},
  \bibinfo{pages}{161107(R)} (\bibinfo{year}{2012}).

\bibitem[{\citenamefont{You, Jian and Wen}(2012)}]{YouWen2}
\bibinfo{author}{\bibfnamefont{Y.-Z.}~\bibnamefont{You}},
\bibinfo{author}{\bibnamefont{C.-M.}~\bibnamefont{Jian}} \bibnamefont{and}
  \bibinfo{author}{\bibfnamefont{X.-G.}~\bibnamefont{Wen}},
  \bibinfo{journal}{Phys. Rev. B} \textbf{\bibinfo{volume}{87}},
  \bibinfo{pages}{045106} (\bibinfo{year}{2013}).
  
\bibitem[{\citenamefont{Barkeshli and Wen}(2010)}]{BarkeshliWen}
\bibinfo{author}{\bibfnamefont{M.} \bibnamefont{Barkeshli}} \bibnamefont{and}
  \bibinfo{author}{\bibfnamefont{X.-G.}~\bibnamefont{Wen}},
  \bibinfo{journal}{Phys. Rev. B} \textbf{\bibinfo{volume}{81}},
  \bibinfo{pages}{045323} (\bibinfo{year}{2010}).

\bibitem[{\citenamefont{Barkeshli and Qi}(2010)}]{BarkeshliQi}
\bibinfo{author}{\bibfnamefont{M.}~\bibnamefont{Barkeshli}} \bibnamefont{and}
  \bibinfo{author}{\bibfnamefont{X.-L.}~\bibnamefont{Qi}},
  \bibinfo{journal}{Phys. Rev. X} \textbf{\bibinfo{volume}{2}},
  \bibinfo{pages}{031013} (\bibinfo{year}{2012}).

\bibitem[{\citenamefont{Barkeshli et~al.}(2012)\citenamefont{Barkeshli, Jian,
  and Qi}}]{Barkeshli}
\bibinfo{author}{\bibfnamefont{M.}~\bibnamefont{Barkeshli}},
  \bibinfo{author}{\bibfnamefont{C.-M.} \bibnamefont{Jian}}, \bibnamefont{and}
  \bibinfo{author}{\bibfnamefont{X.-L.} \bibnamefont{Qi}},
  \bibinfo{journal}{Phys. Rev. B} \textbf{\bibinfo{volume}{87}},
  \bibinfo{pages}{045130} (\bibinfo{year}{2013}).
  
\bibitem[{\citenamefont{Kitaev and Preskill}(2006)}]{Preskill}
\bibinfo{author}{\bibfnamefont{A.~Y.} \bibnamefont{Kitaev}} \bibnamefont{and}
  \bibinfo{author}{\bibfnamefont{J.}~\bibnamefont{Preskill}},
  \bibinfo{journal}{Phys. Rev. Lett.} \textbf{\bibinfo{volume}{96}},
  \bibinfo{pages}{110404} (\bibinfo{year}{2006}).

\bibitem[{\citenamefont{Levin and Wen}(2006)}]{Levin}
\bibinfo{author}{\bibfnamefont{M.}~\bibnamefont{Levin}} \bibnamefont{and}
  \bibinfo{author}{\bibfnamefont{X.-G.} \bibnamefont{Wen}},
  \bibinfo{journal}{Phys. Rev. Lett.} \textbf{\bibinfo{volume}{96}},
  \bibinfo{pages}{110405} (\bibinfo{year}{2006}).

\bibitem[{\citenamefont{Kitaev}(2006)}]{KitaevHoney}
\bibinfo{author}{\bibfnamefont{A.}~\bibnamefont{Kitaev}},
  \bibinfo{journal}{Ann. Phys.} \textbf{\bibinfo{volume}{321}},
  \bibinfo{pages}{2} (\bibinfo{year}{2006}).

\bibitem[{\citenamefont{Wolf et~al.}(2008)\citenamefont{Wolf, Verstraete,
  Hastings, and Cirac}}]{Wolf}
\bibinfo{author}{\bibfnamefont{M.~M.} \bibnamefont{Wolf}},
  \bibinfo{author}{\bibfnamefont{F.}~\bibnamefont{Verstraete}},
  \bibinfo{author}{\bibfnamefont{M.~B.} \bibnamefont{Hastings}},
  \bibnamefont{and} \bibinfo{author}{\bibfnamefont{J.~I.} \bibnamefont{Cirac}},
  \bibinfo{journal}{Phys. Rev. Lett.} \textbf{\bibinfo{volume}{100}},
  \bibinfo{pages}{070502} (\bibinfo{year}{2008}).

\bibitem[{\citenamefont{Eisert et~al.}(2010)\citenamefont{Eisert, Cramer, and
  Plenio}}]{Eisert}
\bibinfo{author}{\bibfnamefont{J.}~\bibnamefont{Eisert}},
  \bibinfo{author}{\bibfnamefont{M.}~\bibnamefont{Cramer}}, \bibnamefont{and}
  \bibinfo{author}{\bibfnamefont{M.~B.} \bibnamefont{Plenio}},
  \bibinfo{journal}{Rev. Mod. Phys.} \textbf{\bibinfo{volume}{82}},
  \bibinfo{pages}{277} (\bibinfo{year}{2010}).

\bibitem[{\citenamefont{Zhang et~al.}(2012)\citenamefont{Zhang, Grover, Turner,
  Oshikawa, and Vishwanath}}]{Zhang}
\bibinfo{author}{\bibfnamefont{Y.}~\bibnamefont{Zhang}},
  \bibinfo{author}{\bibfnamefont{T.}~\bibnamefont{Grover}},
  \bibinfo{author}{\bibfnamefont{A.}~\bibnamefont{Turner}},
  \bibinfo{author}{\bibfnamefont{M.}~\bibnamefont{Oshikawa}}, \bibnamefont{and}
  \bibinfo{author}{\bibfnamefont{A.}~\bibnamefont{Vishwanath}},
  \bibinfo{journal}{Phys. Rev. B} \textbf{\bibinfo{volume}{85}},
  \bibinfo{pages}{235151} (\bibinfo{year}{2012}).

\bibitem[{\citenamefont{Dong et~al.}(2008)\citenamefont{Dong, Fradkin, Leigh,
  and Nowling}}]{Dong}
\bibinfo{author}{\bibfnamefont{S.}~\bibnamefont{Dong}},
  \bibinfo{author}{\bibfnamefont{E.}~\bibnamefont{Fradkin}},
  \bibinfo{author}{\bibfnamefont{R.~G.} \bibnamefont{Leigh}}, \bibnamefont{and}
  \bibinfo{author}{\bibfnamefont{S.}~\bibnamefont{Nowling}},
  \bibinfo{journal}{JHEP} \textbf{\bibinfo{volume}{05}}, \bibinfo{pages}{016}
  (\bibinfo{year}{2008}).

\bibitem[{\citenamefont{Wen}(2003)}]{Wenmodel}
\bibinfo{author}{\bibfnamefont{X.-G.} \bibnamefont{Wen}},
  \bibinfo{journal}{Phys. Rev. Lett.} \textbf{\bibinfo{volume}{90}},
  \bibinfo{pages}{016803} (\bibinfo{year}{2003}).

\bibitem[{\citenamefont{Wen}(2004)}]{Wenmanybody}
\bibinfo{author}{\bibfnamefont{X.-G.} \bibnamefont{Wen}},
  \emph{\bibinfo{title}{Quantum Field Theory of Many-Body Systems}}
  (\bibinfo{publisher}{Oxford Graduate Texts}, \bibinfo{year}{2004}).

\bibitem[{\citenamefont{Fattal et~al.}(2004)\citenamefont{Fattal, Cubitt,
  Yamamoto, Bravyi, and Chuang}}]{Fattal}
\bibinfo{author}{\bibfnamefont{D.}~\bibnamefont{Fattal}},
  \bibinfo{author}{\bibfnamefont{T.~S.} \bibnamefont{Cubitt}},
  \bibinfo{author}{\bibfnamefont{Y.}~\bibnamefont{Yamamoto}},
  \bibinfo{author}{\bibfnamefont{S.}~\bibnamefont{Bravyi}}, \bibnamefont{and}
  \bibinfo{author}{\bibfnamefont{I.~L.} \bibnamefont{Chuang}},
  \bibinfo{journal}{arXiv:quant-ph/0406168v1}  (\bibinfo{year}{2004}).

\bibitem[{\citenamefont{Hamma et~al.}(2005)\citenamefont{Hamma, Ionicioiu, and
  Zanardi}}]{Hamma}
\bibinfo{author}{\bibfnamefont{A.}~\bibnamefont{Hamma}},
  \bibinfo{author}{\bibfnamefont{R.}~\bibnamefont{Ionicioiu}},
  \bibnamefont{and} \bibinfo{author}{\bibfnamefont{P.}~\bibnamefont{Zanardi}},
  \bibinfo{journal}{Phys. Rev. A} \textbf{\bibinfo{volume}{71}},
  \bibinfo{pages}{022315} (\bibinfo{year}{2005}).
  
\bibitem{supp} See Supplemental Material for details on the calculation of the von Neumann entropy of a region enclosing a twist.

\bibitem[{\citenamefont{Vaezi}(2012)}]{Vaezi}
\bibinfo{author}{\bibfnamefont{A.} \bibnamefont{Vaezi}},
  \bibinfo{journal}{Phys. Rev. B} \textbf{\bibinfo{volume}{87}},
  \bibinfo{pages}{035132} (\bibinfo{year}{2013}).

\bibitem[{\citenamefont{Flammia et~al.}(2009)\citenamefont{Flammia, Hamma, Hughes, and Wen}}]{Flammia}
\bibinfo{author}{\bibfnamefont{S.}~\bibnamefont{Flammia}},
  \bibinfo{author}{\bibfnamefont{A.}~\bibnamefont{Hamma}},
  \bibinfo{author}{\bibfnamefont{T.}~\bibnamefont{Hughes}}, \bibnamefont{and}
  \bibinfo{author}{\bibfnamefont{X.-G.} \bibnamefont{Wen}},
  \bibinfo{journal}{Phys. Rev. Lett.} \textbf{\bibinfo{volume}{103}},
  \bibinfo{pages}{261601} (\bibinfo{year}{2009}).

\bibitem[{\citenamefont{Abanin and Demler}(2012)}]{Demler}
\bibinfo{author}{\bibfnamefont{D.}~\bibnamefont{Abanin}} \bibnamefont{and}
  \bibinfo{author}{\bibfnamefont{E.}~\bibnamefont{Demler}},
  \bibinfo{journal}{Phys. Rev. Lett.} \textbf{\bibinfo{volume}{109}},
  \bibinfo{pages}{020504} (\bibinfo{year}{2012}).


  

\end{thebibliography}
\end{document}